\def\na{\nabla}
\def\al{\alpha}
\def\ep{\epsilon}
\def\om{\omega}
\def\sig{\sigma}
\def\p{\partial}
\def\g{{\bf g}}
\def\u{{\bf u}}
\def\n{{\bf n}}
\def\nphi{{\bf N}_\phi}
\def\K{{\bf K}}
\def\OO{{\cal O}}
\def\beq{\begin{equation}}
\def\eeq{\end{equation}}
\def\beqa{\begin{eqnarray}}
\def\eeqa{\end{eqnarray}}

\documentclass{jafm}
\received_date{------}
\accepted_date{------}
\begin{document}
\title{Sessile drop on oscillating incline}
\author{L. De Maio \and F. Dunlop$^{\dagger}$ }
\institute{Laboratoire de Physique Th\'eorique et Mod\'elisation, CNRS UMR 8089
\\ Universit\'e de Cergy-Pontoise, 95302 Cergy-Pontoise, France}
\email{francois.dunlop@u-cergy.fr}
\abstract{Natural or industrial flows of a fluid often involve droplets or 
bubbles of another fluid, pinned by physical or chemical impurities or by the 
roughness of the bounding walls. Here we study numerically one drop pinned on a
circular hydrophilic patch, on an oscillating incline whose angle is 
proportional to $\sin(\omega\,t)$. The resulting deformation of the drop is 
measured by the displacement of its center of mass, which behaves similarly to a
driven over-damped linear oscillator with amplitude $A(\omega)$ and phase lag 
$\varphi(\omega)$. The phase lag is ${\cal O}(\omega)$ at small $\omega$ like a 
linear oscillator, but the amplitude is ${\cal O}(\omega^{-1})$ in a wide range
of large $\omega$ instead of ${\cal O}(\omega^{-2})$ for a linear oscillator. 
 A heuristic explanation is given for this behaviour. The simulations were
performed with the software {\sl Comsol} in mode {\sl Laminar Two-Phase Flow, 
Level Set}, with fluid 1 as engine oil and fluid 2 as water.}
\maketitle

\section{Introduction}
Equilibrium of a drop pinned on an incline was studied by many authors, see 
\cite{DDH17} and references therein. Shape and motion of drops sliding
down an inclined plane have also been studied, see 
\cite{LDL05} and references therein. 
 Drops on vibrating horizontal surfaces have been the subject of much 
interest recently from experimental, theoretical or numerical points of view. 
The vibrations or oscillations of the substrate can be horizontal
\cite{DCG05,LLS04,DCC06,CK06},
or vertical
\cite{LLS06}.
The effect of vibrations on hysteresis, pinning and depinning, was studied in 
particular by \cite{NBB04,VSG07}.
The effect of vibrations on the Cassie-Wenzel transition was studied in 
particular by \cite{BC09,BPWE07}.
A review of drop oscillations is given by \cite{MDCA14}. 
More recent experimental results and references are found in \cite{RE17}.

Here we consider the case of an oscillating
incline where the angle $\al(t)$ of the slope follows
\beq\label{alphat}
\al(t)={\pi\over4}\sin(\om\, t)
\eeq
while the circular basis of the drop remains fixed on the incline as a disc of
radius $r$. We keep the frame of reference attached to the incline, so that the 
gravity vector oscillates:
\beq
\g=\left(\begin{matrix}
g\sin(\al(t))\cr
0\cr
-g\cos(\al(t))\end{matrix}\right)
\eeq
We assume that the inertial pseudo-forces per unit volume, like the centrifugal
force, are negligible with respect to gravity, which will be the case if
$\om^2\,r\ll g$.

\begin{figure}
  \centering
  \includegraphics[width=\columnwidth]{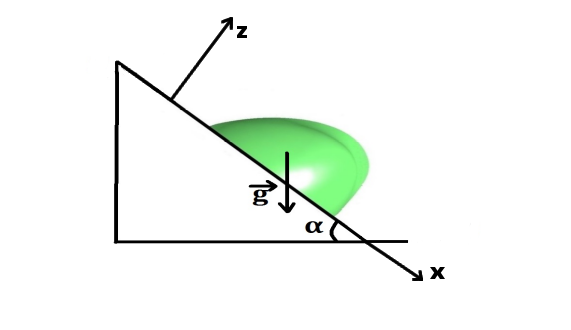}\\ 
\caption[]{ Water drop at equilibrium pinned on incline of angle 
$\al=\pi/4$. Bond number $Bo=0.22$.  The drop is surrounded by oil.}
  \label{fig:iso}
\end{figure}

The Bond number is the ratio between gravity and capillarity, and we define it 
precisely as
\beq
Bo={\Delta\rho g r^2\over\sigma}
\eeq
where $\Delta\rho$ is the density difference between the two fluids and $\sigma$
is the interface tension. We are interested in moderate but significant drop 
deformations, with Bond number of order one, as shown on Fig. \ref{fig:iso}. All
the simulations presented here will be with $Bo=0.22$.
For bond number of order one and pulsation $\om$ not much larger than the 
natural pulsation of the drop, the fluid velocity will vary from 0 to about
$\om r$ over a distance $r$. This motivates a Reynolds number defined as
\beq
Re_\om={\rho\om\, r^2\over\eta}
\eeq
with $\rho=\rho_{\rm water}$ and $\eta=\eta_{\rm water}$. 
In the same regime the  quadratic term in the Navier-Stokes equation 
(\ref{NS}) will be of order $\rho\om^2\,r$, the same as the centrifugal force 
per unit volume. Therefore it will be consistent,
 and will save some computing time, to neglect it (Stokes flow). 
 The equation remains non-linear due to the interfacial tension force. 

For pulsations $\om$ larger than the 
natural pulsation of the drop, the response of the drop and the actual velocity
will be much smaller. A Reynolds number using the maximum measured velocity
will always be less than 1 in our simulations. Viscosity plays an essential role
in the present study, which does not allow short-cuts such as interface motion 
by curvature based on the Laplace-Young equation.
\section{Diffuse interface and level set method}
The sharp interface between immiscible fluids is replaced by a diffuse interface
spreading over a few mesh elements across the physical interface. A level set 
function $\phi$, inspired by van der Waals, goes smoothly from zero to one when 
crossing the interface from fluid 1 into fluid 2. The mixture obeys the 
Navier-Stokes equation for an incompressible fluid,
\beqa
\rho{\p\u\over\p t}+\rho(\u\cdot\na)\u&=&\na\cdot\Bigl[-p{\bf I}
+\eta\bigl(\na\u+\na\u^T\bigr)\Bigr]\cr
&&\hskip1.5cm+\rho{\bf g}+{\bf f}_{st}\label{NS}\\
\na\cdot\u&=&0
\eeqa
where ${\bf I}$ is the identity matrix,
and we neglect the quadratic term in (\ref{NS}). The density and dynamic 
viscosity are functions defined by
\beqa
\rho=(1-\phi)\rho_1+\phi\rho_2\cr
\eta=(1-\phi)\eta_1+\phi\eta_2
\eeqa
The surface tension force per unit volume ${\bf f}_{st}$ is
\beq\label{st}
{\bf f}_{st}=\na\cdot\Bigl(\sigma\bigl({\bf I}-(\nphi\nphi^T)\bigr)\delta\Bigr)
\eeq
where $\sigma$ is the interfacial tension, $\nphi=\na\phi/|\na\phi|$ is a normal
vector also defined in the bulk, and $\delta=6|\na\phi|\,\phi(1-\phi)$ is a 
smooth Dirac delta function concentrated near the interface, which is the level 
set $\{\phi=0.5\}$. Formula (\ref{st}), being the divergence of a flux, can be
integrated by parts in the weak form of the partial differential equation, and then requires just one derivative of $\phi$. It was 
shown by \cite{LNSZ94} to be a smooth approximation to the usual Laplace force 
$\sigma H\nphi\delta({\rm interface})$ where $H$ is the mean curvature of the 
interface and $\delta({\rm interface})$ is a true Dirac delta function supported
by the interface.

The level set function $\phi$ obeys
\beq\label{LS}
{\p\phi\over\p t}+\u\cdot\na\phi=\gamma\na\cdot
\Bigl(\ep\na\phi-\phi(1-\phi){\na\phi\over|\na\phi|}\Bigr)
\eeq
where $\ep$ in the diffusion term controls the interface thickness. It will be 
taken as $h/2$, half the mesh size. The parameter $\gamma$ is a constant with 
the dimension of a velocity, which we fix as $r\,\om/(2\pi)$ where $r$ is the 
initial radius of the drop. The level set method for two phase flow was 
developed in particular by \cite{OK05}.
\begin{figure}
\centering\includegraphics[width=\columnwidth]{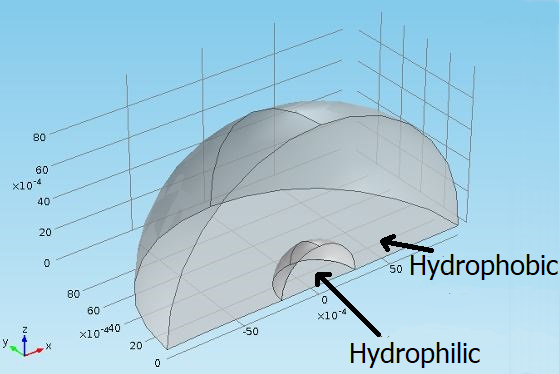}\\
\caption[]{Setup.}\label{fig:diapo3}
\end{figure}
\section{Setup}
The incline is designed with a circular hydrophilic patch of radius $r=2.5\,$mm
and the remaining surface hydrophobic. The corresponding Young contact angles 
are set to 0 degree (perfectly hydrophilic) and 180 degrees (perfectly 
hydrophobic) respectively. 

A water drop of volume $2\pi r^3/3$ is deposited on the hydrophilic patch.
 The vessel is filled with oil, and closed with no air inside. 
In the absence of gravity, the drop is a hemisphere, with contact angle $\pi/2$.
This will also be the initial configuration in our simulations.

The vessel is intended to be large with respect to the water drop, so that
friction occurs only near the drop. The Archimedes force, encapsulated in the 
pressure and gravity terms of the Navier-Stokes equation, does not depend upon
the volume of the vessel. For simulation purposes, we have to use a simulation
box of modest size. The effect of the box will be minimized if it has the
symmetry of the problem at lowest order, hence a hemisphere with same center
as the initial drop, and we choose its radius as four times the initial drop
radius. On it we choose ``slip'' boundary conditions:
impenetrable and frictionless, again to minimize the effect of having a 
relatively small simulation box. 

The center of the hydrophilic patch is chosen as origin of coordinates and the 
$z$-axis perpendicular to the incline. The incline then starts 
oscillating around the $y$-axis according to (\ref{alphat}).
The plane $\{y=0\}$ is a plane of symmetry,  allowing to make the study in a
quarter of a sphere, see Fig.~\ref{fig:diapo3}. 

In the stationary regime, the contact angles at the front ($x=r$) and at the 
back ($x=-r$) will oscillate between a minimum angle $\theta^{\rm min}$ and a 
maximum angle $\theta^{\rm max}$. So long as the maximum contact angle remains 
strictly less than 180 degrees, the contact line cannot move into the 
hydrophobic region.  
So long as the minimum contact angle remains strictly larger than 0 degree, 
the contact line cannot move into the hydrophilic region. 
The role of the substrate is to ensure pinning.

For real substrates, the advancing angle $\theta^A$ of the hydrophobic material 
and the receding angle $\theta^R$ of the hydrophilic material will replace 180 
degrees and zero degree respectively. The scope of our study is bounded by the 
conditions
\beq
0\le\theta^R<\theta^{\rm min}<\theta^{\rm max}<\theta^A\le\pi\,.
\eeq
It implies bounds on Bond number and slope angle $\al$, which are satisfied
in the present study. It would be interesting to go beyond and also study 
depinning. This is left to future work.
\begin{center}
\begin{tabular}{|c|c|c|c|} \hline
 \multicolumn{4}{|c|}{\bf Table 1: Physical parameters}  \\ \hline
 & $\rho\,$[kg/m$^3$] & $\eta\,$[Pa\,s] & $\sigma\,$[N/m] \\ \hline
Engine oil  & 888  & 0.079  &   \\ \hline
Water  & 1000  & 0.001  &   \\ \hline
Interface  &   &   & 0.031  \\ \hline
\end{tabular}
\end{center}
\section{{\sl Comsol}}
We used the finite elements software {\sl Comsol} (see https://www.comsol.com/)
in mode {\sl Laminar Two-Phase Flow, Level Set}, with fluid 1 as engine oil and
fluid 2 as water, at 20$^\circ$C, see Table 1. 
\begin{figure}
\centering\includegraphics[width=\columnwidth]{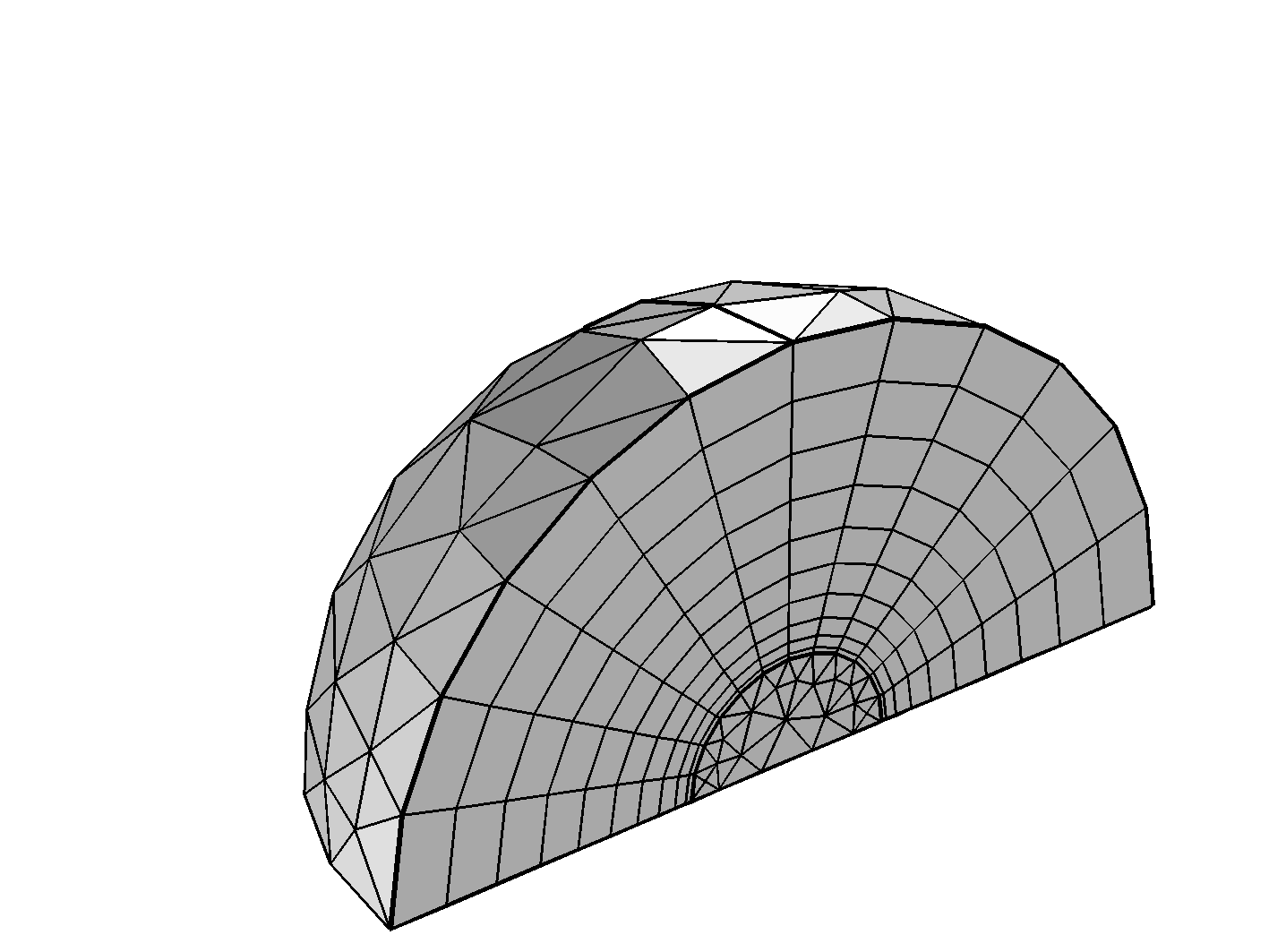}\\
\caption[]{Mesh.}\label{fig:mesh}
\end{figure}
The simulation box is a quarter of a sphere of radius $4r$. The outer sphere is
not a physical boundary, and on it we choose {\sl slip} boundary conditions:

\beqa
\u\cdot\n=0&\cr
\K=(\K\cdot\n)\cdot\n\,,&\quad \K=\eta\bigl(\na\u+\na\u^T\bigr)\n\hskip0.5cm 
\eeqa
where $K$ is the viscous stress vector upon an infinitesimal surface of normal 
$\n$.
The symmetry plane $\{y=0\}$ obeys the same boundary conditions, with also
\beq
\na\cdot\nphi=0
\eeq
The hydrophilic patch is a {\sl wetted wall} with contact angle $\theta_w=0$,
meaning a boundary condition
\beq
\sig(\n-\nphi\cos\theta_W)\delta={\eta\over\beta}\u
\eeq
where $\beta$ is a slip length equal to the mesh size $h$. The remaining part of
the incline is a {\sl wetted wall} with contact angle $\theta_w=\pi$.
The mesh is built as shown on Fig. \ref{fig:mesh}, with maximal mesh size 
$h=0.4\,$mm and $h\sim 0.1\,$mm in the region of the interface, leading to 24560
degrees of freedom. 

We impose at least one time step in every 1/40 of a period so as to be able to 
distinguish a sinusoidal response. With the chosen mesh, it turns out that 
{\sl Comsol} does not need smaller time steps to satisfy its default tolerance.
Each run for one value of $\om$ took about 20 hours with an Intel 
i7-3770\,CPU@3.40GHz x8. 
\begin{figure}
\centering\includegraphics[width=\columnwidth]{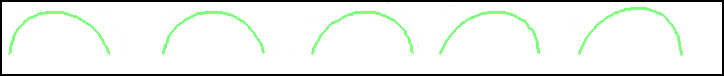}\\
\caption[]{Trace of the drop on the symmetry plane $\{y=0\}$ at five times. 
$Bo=0.22$, $\om=0.1\,$s$^{-1}$.}
\label{fig:film}
\end{figure}
\section{Results}

The trace of the drop on the symmetry plane $\{y=0\}$ at different times is 
shown on Fig. \ref{fig:film}.  

\begin{figure}
  \centering  \includegraphics[width=\columnwidth]{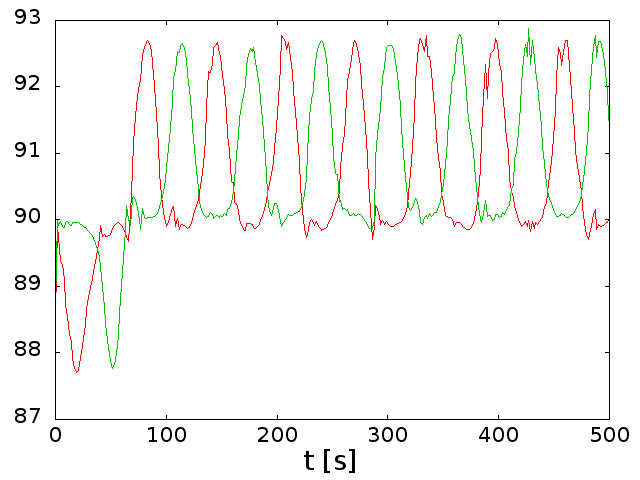}\\
\caption[]{Contact angles $\theta^r(t)$ at the front (red) and $\theta^{-r}(t)$
at the back (green), measured in degrees as (\ref{nnphi}), for $Bo=0.22$, 
$\om=0.1\,$s$^{-1}$.}  \label{fig:thetarl}
\end{figure}

\begin{figure}
  \centering  \includegraphics[width=\columnwidth]{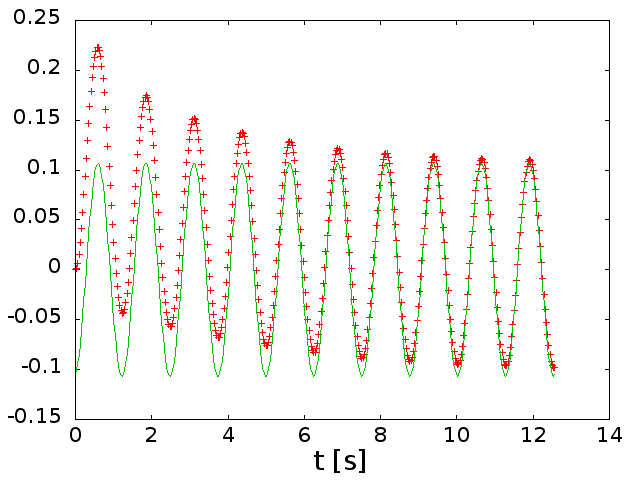}\\
  \caption[]{Normalized abscissa of water center of mass $\bar x(t)$, as
Eq. (\ref{xt}) (red +), and sinusoidal fit of permanent 
regime, $A\,\sin(\om\,t-\varphi)$ as (\ref{sin}) (green, continuous), for $Bo=0.22$, $\om=5\,$s$^{-1}$.}
  \label{fig:om5x10T}
\end{figure}

 The contact angles $\theta^r(t)$ and $\theta^{-r}(t)$ at the front and the 
back are shown on Fig. \ref{fig:thetarl}. These contact angles are measured as
\beq\label{nnphi}
\theta=\arccos(\n\cdot\nphi)
\eeq
at $(x,y,z)=(r,0,0)$ and $(x,y,z)=(-r,0,0)$ respectively. When the incline is 
set in motion, at $t=0$, the liquid drop does not follow instantaneously, whence
a start below 90 degrees. In the stationary regime a noticeable feature is that 
the contact line spends more time near the minimum than near the maximum. 

Eq. (\ref{nnphi}) is a measurement of the interface normal, pointing from oil 
into water, at a single point, which is a mesh vertex. Small numerical errors 
are clearly visible in Fig. \ref{fig:thetarl}. A systematic error is also 
present: when the contact angle 
approaches $\theta^{\rm max}$, the contact line goes slightly into the hydrophobic
region. Similarly when the contact angle approaches $\theta^{\rm min}$, the 
contact line goes slightly into the hydrophilic region. Measuring the contact 
angles at $x=\pm r$ underestimates the amplitude of oscillations.  

Measuring contact angles, experimentally or numerically, is subject to debate,
especially in dynamics. Fitting individual images of a film is tedious and 
systematic deviations may also be present if the fit is over a length where 
gravity produces bending. In dynamics the bending effect of gravity cannot be 
computed exactly. We have therefore chosen to analyse the data in terms 
of the motion of the centre of mass of the drop, whose definition is
obvious and whose statistics is optimal.

\begin{figure}
\centering\includegraphics[width=\columnwidth]{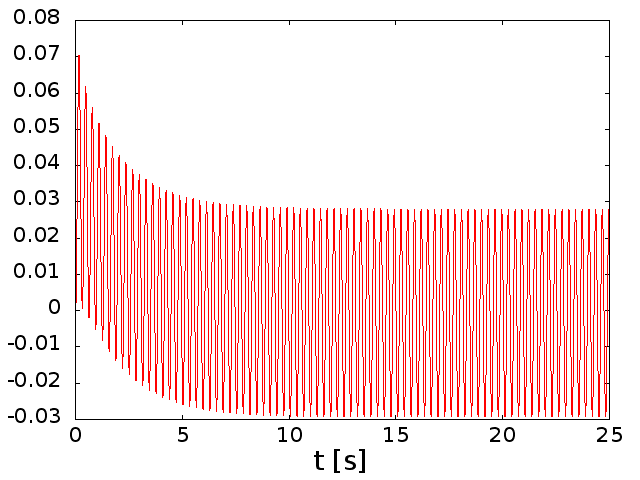}\\
\caption[]{Normalized abscissa of water center of mass $\bar x(t)$, 
Eq. (\ref{xt}), for $\om=20\,$s$^{-1}$.}
\label{fig:om20x}
\end{figure}

The abscissa of the center of mass of water is recorded, normalized arbitrarily
using the drop basis radius $r$ and the volume $\pi r^3/3$ of a quarter of a 
sphere of radius $r$\,:  
\beq\label{xt}
\bar x(t)={\int dxdydz\ \phi(x,y,z,t)\,x\over\pi r^4/3}
\eeq
The integral is over the simulation box, namely a quarter of a sphere of radius 
$4r$. After a transient, which lasts longer for larger $\om$, the system 
approaches a stable permanent regime, as shown on Figs. \ref{fig:om5x10T},
\ref{fig:om20x}. The finite elements method does not conserve exactly the total 
mass of each fluid, and a small parasitic drift is often present in simulations,
but it is not the case here.

We then use the {\sl gnuplot} fit, a nonlinear least-squares 
Marquardt-Levenberg algorithm, and search for an amplitude $A$ and a phase lag
$\varphi$ such that  
\beq\label{sin}
\bar x(t)-A\,\sin(\om\,t-\varphi)\,\longrightarrow0\quad{\rm as}\quad 
t\to\infty
\eeq
Results are like the example shown on Fig. \ref{fig:om5x10T}, where the error
measured by the $rms$ of residuals over one period falls below 1\% after a few 
periods (after 7 periods in the example shown). The resulting incertainties over
$A$ and $\varphi$ are also below 1\%.

A sinusoidal response with the same $\om$ as the incline angle was to be
expected for a linear system. We used the Stokes equation with a non-linear
surface tension force, and the transport equation (\ref{NS}) is also non-linear.
 
\begin{figure}
\centering\includegraphics[width=\columnwidth]{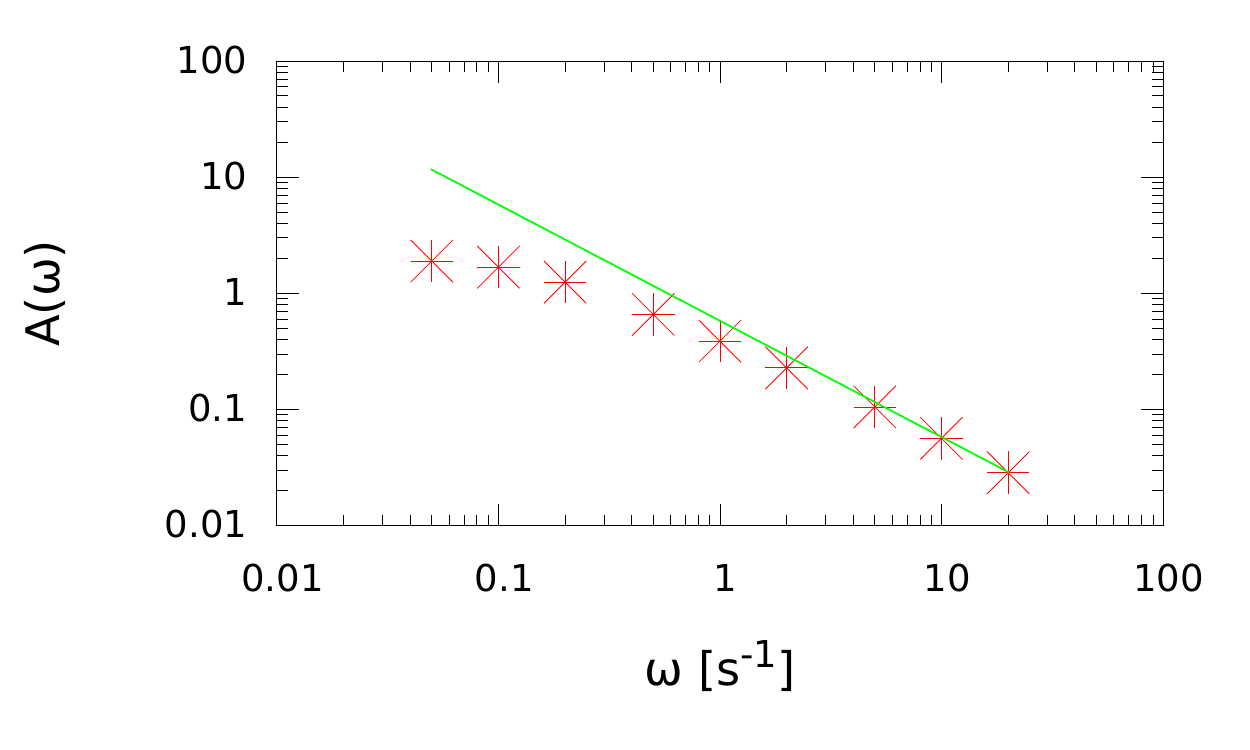}\\
\caption[]{Amplitude $A(\om)$ from (\ref{sin}) with asymptote $0.58/\om$.}
\label{fig:xphi}
\end{figure}

\begin{figure}
\centering\includegraphics[width=\columnwidth]{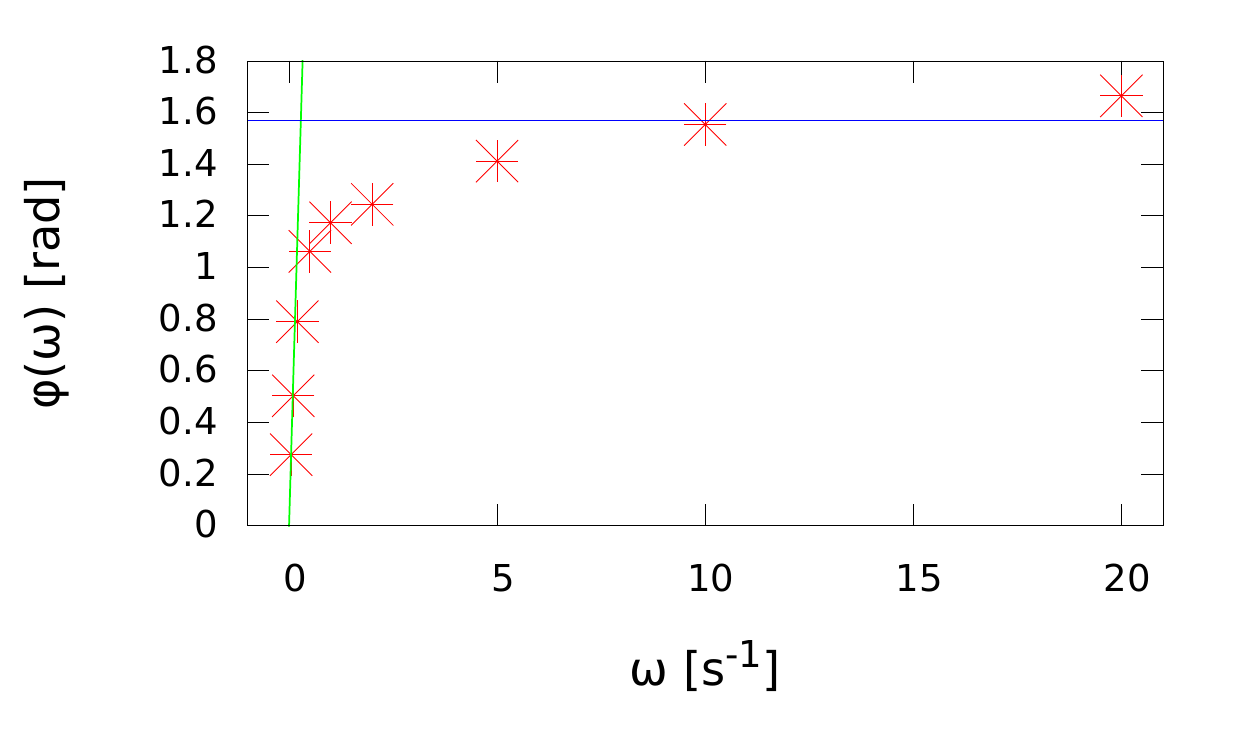}\\
\caption[]{Phase lag $\varphi(\om)$ from (\ref{sin}) with line at $\pi/2$ and 
tangent at the origin $\varphi=5.5\om$.}
\label{fig:xphiphi}
\end{figure}

Results are listed in Table 2, where the case $\om=0$ is in fact the limit as 
$\om\to0$, namely the stationary case $\al(t)=\pi/4\ \forall t>0$. Plots of $A$
and $\varphi$ versus $\om$ are given in Fig. \ref{fig:xphi} and
Fig. \ref{fig:xphiphi}. They look much like a driven over-damped linear 
oscillator, with a notable exception: the amplitude of the permanent 
oscillations behaves like $\om^{-1}$ at large $\om$ instead of $\om^{-2}$ for the 
driven damped linear oscillator. The phase lag $\varphi(\om)$ is proportional
to $\om$ at small $\om$  like a driven damped linear oscillator. 
\begin{center}
\begin{tabular}{|c|c|c|} \hline
 \multicolumn{3}{|c|}{\bf Table 2: results}  \\ \hline
$\om\,$[s$^{-1}$] & $A$ & $\varphi\,$[rad]  \\ \hline
0  & 1.92  & 0   \\ \hline
0.05  & 1.89 & 0.275    \\ \hline
0.1  & 1.68   & 0.503   \\ \hline
0.2  & 1.25  &  0.79  \\ \hline
0.5  & 0.66  &  1.06  \\ \hline
1  & 0.387  &  1.17  \\ \hline
2  & 0.228  & 1.24    \\ \hline
5  & 0.105  & 1.41   \\ \hline 
10  & 0.0564  & 1.55   \\ \hline 
20  & 0.0286  & 1.66   \\ \hline 
\end{tabular}
\end{center}

\section{Heuristics for $\om\to\infty$.}
Let us first review the case of a driven solid oscillator, subject to a fluid
friction force, obeying a differential equation of the form
\beq\label{osc2}
\ddot x + {\rm friction\ force} + {\rm restoring\ force}=\sin\om\,t
\eeq 
As $\om\to\infty$ we don't expect resonance. Therefore each term on the 
left-hand-side will be of order at most the order of the right-hand-side,
namely $\OO(1)$. We expect a periodic permanent regime of period $T=2\pi/\om$.
If the amplitude is $A$ then $\ddot x$ will be of order $A\om^2$, implying
$A$ of order at most $\om^{-2}$. The restoring force will be $o(1)$, the velocity
of order $A\om\sim\om^{-1}$ and the fluid friction force $o(1)$.
Therefore, as $\om\to\infty$, the system tends to $\ddot x=\sin\om\,t$,
leading to an amplitude $\om^{-2}$ and phase lag $\pi$, in agreement with the 
exact solution of the linear case.

Another driven system may obey a first order differential equation of the form
\beq\label{osc1}
\dot x + {\rm restoring\ force}=\sin\om\,t
\eeq 
Again we expect a periodic permanent regime of period $T=2\pi/\om$,
and no resonance, so that each term on the left-hand-side will be of order at 
most $\OO(1)$. If the amplitude is $A$ then $\dot x$ will be of order $A\om$, 
implying $A$ of order at most $\om^{-1}$. The restoring force will be $o(1)$. 
Therefore, as $\om\to\infty$, the system tends to $\dot x=\sin\om\,t$,
leading to an amplitude $\om^{-1}$ and phase lag $\pi/2$, in agreement with the 
exact solution of the linear case.

We have studied a drop on an oscillating incline in a regime where the inertial
forces, such as the centrifugal force, are negligible with respect to gravity
and capillarity, both of same order for Bond number of order one.
We thus have $\om^2r\ll g$. The acceleration term $\p u/\p t$ in the 
Navier-Stokes equation is of same order and therefore negligible.
And the Reynolds number was always less than one so that the quadratic term in 
the Navier-Stokes equation could be neglected. Therefore a behaviour 
corresponding to (\ref{osc1}) rather than (\ref{osc2}) should be observed, 
leading to an amplitude $A\sim\om^{-1}$ rather than $A\sim\om^{-2}$ as 
$\om\to\infty$.

In simple words: a liquid drop can deform in many 
different ways and will do so as far as the shear stress $\na \u$ remains 
bounded. If $A$ is the amplitude of the motion of the center of mass in the
frame of reference of the incline, then $\na \u$ is of order $A\om/r$,
giving $A\sim\om^{-1}$ as $\om\to\infty$.

When $\om\to0$, acceleration is negligible in all cases, and both solid and 
liquid oscillators have an amplitude $\OO(1)$ and a phase lag $\OO(\om)$.

\section{Conclusion}
A sessile millimetric droplet on an incline responds similarly to a driven 
damped linear oscillator to a sinusoidal oscillation of the angle of the 
incline. However, the amplitude of the drop deformation is proportional to
$\om^{-1}$ at large $\om$, whereas a simple pendulum on an oscillating incline 
responds with an amplitude proportional to $\om^{-2}$ at large $\om$.

The diffuse interface modelisation imply diffusion times larger than the true
physical times, but the discrepancy should go to zero with finer and finer
meshes. Also, because there is more space for water (originally in the small 
sphere) to diffuse into oil (originally in the large sphere), than conversely, 
the level set 0.5, considered as the interface, shrinks a little during the 
first seconds. This effect should also go to zero with finer and finer
meshes.

Beyond $\om\sim20\,$s$^{-1}$, in the oscillating frame of reference,  one cannot 
neglect the inertial pseudo-forces. One can expect that including the 
centrifugal force in the Navier-Stokes equation would increase the drop 
deformation at large $\om$.
\bibliography{ldfd}
\end{document}